# Variability as a Predictor: A Bayesian Variability Model for Small Samples and Few Repeated Measures


Joshua F. Wiley[1, 2]

Bei Bei[3, 4]

John Trinder[4]

Rachel Manber[5]

1. Elkhart Group Ltd., Columbia City, IN, USA

2. Department of Psychology, University of California Los Angeles, Los Angeles, CA, USA

3. School of Psychological Sciences, Faculty of Biomedical and Psychological Sciences, Monash University, Victoria, Australia

4. Melbourne School of Psychological Sciences, University of Melbourne, Victoria, Australia

5. Stanford University School of Medicine, Department of Psychiatry and Behavioral Science Stanford University, CA, USA


## Author Note


The authors wish to acknowledge the adolescents who volunteered their time to participate in the sleep study.

Wiley owns shares in Elkhart Group Ltd.



Correspondence concerning this article should be addressed to

Attn: Joshua F. Wiley

1873 E. Ravenwood Ln.,

Columbia City, Indiana, 46725

Phone: +1 260 673 5518

Email: josh@elkhartgroup.com





**Abstract**

Whilst most psychological research focuses on differences in means, a growing body of literature demonstrates the value of considering differences in *intra*-individual variability. Compared to the number of methods available for analyzing mean differences, there is a paucity of methods available for analyzing intra-individual variability, particularly when variability is treated as a predictor. In the present article, we first reviewed methods of analyzing intra-individual variability as an outcome, including the individual standard deviation (*ISD*) and some recent methods. We then introduced a novel Bayesian method for analyzing intra-individual variability as a predictor. To make this method easily accessible to the research community, we developed an open source R package, VARIAN. To compare the accuracy of parameter estimates using the proposed Bayesian analysis against the *ISD* as a predictor in a regression, we carried out a simulation study. We then demonstrated, using empirical data, how the estimated intra-individual variability derived from the proposed Bayesian analysis can be used to answer the following two questions: (1) is intra-individual variability in daily time-in-bed associated with subjective sleep quality? (2) does subjective sleep quality mediate the association between time-in-bed variability and depressive symptoms? We concluded with a discussion of methodological and practical considerations that can help guide researchers in choosing methods for evaluating intra-individual variability.

The majority of psychological research examines mean (also known as "location") differences, such as mean differences between groups, how means can be predicted by other factors, and how means change over time.  Researchers also have acknowledged from both empirical and theoretical perspectives, that inter- and intra-individual variability (also known as "scale") are important to many areas of psychological research (e.g., Eizenman, Nesselroade, Featherman, & Rowe, 1997).  For example, empirical research by Russell, Moskowitz, Zuroff, Sookman, and Paris (2007) demonstrated that patients with borderline personality disorder, which is characterized by unstable relationships and affective instability (American Psychiatric Association, 2013), exhibited significantly higher intra-individual variability in affect than did controls.  In the context of aging and developmental processes, Ram and Gerstorf (2009) overviewed important conceptual, methodological, and research design considerations for studying intra-individual variability.  Conceptual and empirical interest in scale is also growing in the field of sleep (e.g., Buysse et al., 2010; Suh et al., 2012).  Thus interest is growing in studying and understanding the impact of intra-individual variability in multiple fields of psychological inquiry.

The aim of this article is to provide researchers with a rigorous and accessible method for estimating intra-individual variability, and highlight the utility of the estimated index of variability when intra-individual differences are of theoretical and empirical relevance.  We first review existing methods for quantifying and modeling variability, including examples of commonly used approaches and recent methodological advances.  Next we introduce a flexible and general Bayesian variability model for using variability as a predictor.  An open-source R package "varian" was developed for easy implementation of this method.  We present a simulation study that investigates the performance of the proposed Bayesian method in



conditions that vary by sample size, number of repeated measures, effect size, and individual

differences in variability. Findings are contrasted with results using individual standard

deviations (*ISD*s), an earlier approach to measuring variability. The application of the proposed

Bayesian method is then demonstrated through two empirical examples, each includes sample

code, model diagnostics, and interpretation of results. Finally, recommendations are made for

how to select a suitable method for variability analysis, and extensions to the proposed method

are discussed.

## Methods for Quantifying and Modeling Variability

One common measure of variability is the standard deviation. With this method, intra-

individual variability is calculated as the standard deviation of observations for each subject, also

known as individual standard deviation (*ISD*). The *ISD* may subsequently be used as either an

outcome or a predictor. As the *ISD* is based on deviations from individuals' means, systematic

time effects (e.g., a linear increase over time) are incorporated into the computation of the *ISD*,

and as a result, may provide an inflated or otherwise biased estimation of intra-individual

variability. For example, while sleep researchers are interested in night-to-night variability of

sleep schedules, and sleep clinicians often recommend alterations to sleep timing and duration,

other factors, such as seasonal variation in daylight, can affect individuals' sleep/wake

behaviours, and such variations may not be of direct research or clinical relevance. In these

cases, the *ISD* may overestimate the type of intra-individual variability that is of interest. Such

biases can be addressed by calculating the *ISD* on the residuals, after adjusting for time or other

relevant factors (i.e., detrending).

A common method of quantifying variability that is less sensitive to systematic changes

is the root mean square of successive differences (RMSSD; von Neumann, Kent, Bellinson, &



Hart, 1941) . Other measures of variability are: the variance (i.e., $ISD^2$), mean square of successive differences ($RMSSD^2$), median absolute difference, the range, interquartile range, and coefficient of variation. One limitation shared by all of these approaches is that they do not account for measurement error. In contrast to means, which have good reliability even with a few repeated measures, the $ISD$ has poor reliability, particularly when the number of repeated observations is small and when individual difference in $ISD$s are small (Estabrook, Grimm, & Bowles, 2012). With analytically derived reliabilities of $ISD$ and $ISD^2$, as many as 50 repeated observations were required for a reliable questionnaire (reliability 0.9) to estimate intra-individual variability with reasonable reliability (reliability 0.8) (Wang & Grimm, 2012). It is sometimes not feasible to collect these many repeated measures due to participant burden and cost.

Accounting for measurement error is a well-studied topic in psychometrics. Methods, such as structural equation modelling, using latent variables, have been developed to account for measurement error in modeling (e.g., Bollen, 1989). However, most statistical models are designed to test location (mean), not scale (variability) effects. For independent observations, flexible generalized additive models for location and scale have been developed (Rigby & Stasinopoulos, 2005; Stasinopoulos & Rigby, 2007), but these are not applicable to repeated measures data (dependent observations) and thus to the study of intra-individual variability.

Several recent methodological advances facilitate the empirical investigation of variability. These methodologies allow testing of scale effects in non-independent observations by treating variability as a latent variable in mixed effect or multi-level models. Hedeker, Mermelstein, and Demirtas (2008) developed mixed-effects location and scale models in a maximum likelihood framework. This method predicts both between- and within-subject



factors, including random intercepts and random variabilities. It has been applied to analyze

ecological momentary assessment data (for an excellent introduction to the application, see

Hedeker, Mermelstein, & Demirtas, 2012).  Subsequently, Hedeker, Demirtas, and Mermelstein

(2009) extended mixed effects location and scale models from continuous to ordinal data, and Li

and Hedeker (2012) extended them to three-level models.

      Another approach, proposed by Jahng, Wood, and Trull (2008), is to first calculate the

squared successive differences, and then model these using a generalized linear mixed model.

This approach uses successive differences and is therefore less sensitive to systematic intra-

individual changes.  The intercept, effectively the mean squared successive difference

($RMSSD^2$), can be predicted by entering each squared successive difference into a generalized

linear mixed model.  One limitation of this approach is its ability to handle missing data.  For

example, if data are collected across three days, there are two successive differences, $t_2 - t_1$, and

$t_3 - t_2$, and if the second day ($t_2$) is missing, both successive differences are undefined.

      Wang, Hamaker, and Bergeman (2012) proposed a multilevel model that can be

estimated in one step in a Bayesian framework using the raw data, without having to first

calculate the successive differences.  An important advantage of this method is that it models

both temporal dependency (captured via an autocorrelation coefficient) and magnitude of

variability, whereas the Hedeker et al. (2008) method models only the magnitude.  Modeling

temporal dependence may be particularly valuable when the number of repeated measures is

large.  For example, a positive auto-correlation indicates a higher score on one day is associated

with a higher score the next day, whereas a negative auto-correlation indicates a higher score on

one day is associated with a lower score on the next day.  A limitation of modelling magnitude

and temporal dependence is that models with few repeated measures may not converge and thus



fail to estimate variability. Wang et al. (2012) observed this limitation. In a sample of over 200 participants, their method converged when using all 56 repeated measures per participant, but did not converge when using 7 or even 14 repeated measures per participant.

Another advance, which represents an important and valuable alternative to the ones described above, uses differential equations in structural equation models or dynamical systems models (Boker & Nesselroade, 2002) and derivatives with bootstrapping for statistical inference (Deboeck, Montpetit, Bergeman, & Boker, 2009). However, this advance is less relevant to the current paper, which focuses on modeling variability as the scale of a distribution.

In summary, simple methods of quantifying variability such as the *ISD* are convenient but suboptimal, as they have low reliability and do not account for measurement error, resulting in decreased power to detect effects. Other methods, such as those proposed by Hedeker et al. (2008), Jahng et al. (2008), and Wang et al. (2012), represent superior alternatives when researchers are interested in predicting intra-individual variability. The method developed by Hedeker et al. (2008) may be more appropriate when fewer repeated measures are available, at the cost of not modeling temporal dependence. To our knowledge, it is also the only option available for estimating variability of ordinal data.

However, with the exception of the ISD, the existing methods assume that intra-individual variability is the outcome of interest and have not produced an index of intra-individual variability that can then be used as a predictor of other outcomes. In the next section, we introduce a Bayesian variability model that provides estimates of intra-individual variability and uses the estimated index of variability as a *predictor* of an outcome that is reliable with as few as five repeated measures.



## Intra-individual Variability as a Predictor

**Overview**

Conceptually, the proposed model has two parts.  The first part is an extension of a multilevel regression model and estimates intra-individual variability, accounting for measurement error.  The estimate of intra-individual variability represents individuals' variability around their own mean.  The method for estimating intra-individual variability uses a regression framework, and as such it has all standard features of regressions.  For example, it is possible to control for variables, such as time, day of the week, or other relevant variables by including them as covariates.  When a covariate, such as time, is in the model, the estimate of intra-individual variability would be variability around the time trend, rather than around individuals' means (i.e., detrended variability in residuals after controlling for time as a covariate).  The second part of the model uses the estimate of intra-individual variability, accounting for measurement error, as a predictor in a multiple regression predicting another outcome.

For practical reasons the method we propose employs Bayesian estimation through Markov Chain Monte Carlo (MCMC) simulation.  Bayesian MCMC methods provide a flexible framework that allows non-normally distributed outcomes and testing of indirect effects in mediation analysis (Yuan & MacKinnon, 2009).  Bayesian methods have been gaining attention recently in the behavioral sciences (e.g., Kruschke, 2010a; Muthén & Asparouhov, 2012).  For an accessible introduction, see Kruschke (2010b) and Kruschke (2013).

**Bayesian Variability Model**

Let $Y$ be a between-person outcome, and $V$ a within-person variable, both measured repeatedly on each person.  The goal is to use each individual's variability in $V$ to predict their



score on $Y$, and, if desired, estimate each individual's mean of $V$ as an additional predictor of $Y$. Accounting for the effect of the average of $V$ may be particularly important given that at least in some domains, the means and standard deviations are strongly correlated (Nesselroade & Salthouse, 2004). Consider a multilevel model for $V$, for simplicity and without loss of generality, we demonstrate an unconditional model:

$$V_{ij} \sim N(\mu_j, \sigma)$$

where $V_{ij}$ is the value for the $i$th ($i = 1, 2, ..., I_j$) assessment for the $j$th ($j = 1, 2, ..., N$) subject. A unique mean is estimated for each subject, and these are assumed to be normally distributed:

$$\mu_j \sim N(\mu_\mu, \sigma_\mu)$$

The model assumes that each $V_{ij}$ observation comes from a normal distribution with some mean and standard deviation. The individual means ($\mu_j$) are in turn assumed to come from a normal distribution with an overall grand mean ($\mu_\mu$) and standard deviation ($\sigma_\mu$). We extend this model by allowing the standard deviation to also vary by subjects, that is:

$$V_{ij} \sim N(\mu_j, \sigma_j)$$

Where the $\sigma_j$ term represents individual standard deviations or in the case of a conditional model, individual residual standard deviations. These are assumed to come from a gamma distribution with scale and shape parameters $\alpha$ and $\beta$:

$$\sigma_j \sim \Gamma(\alpha, \beta)$$

Estimates of the individual standard deviations, $\sigma_j$, then are used as predictors of the outcome, $Y$, using a standard multiple linear regression, that is:

$$Yj \sim N(\mu_{2j}, \sigma_2)$$

where

$$\mu_{2j} = \beta_1 + ... + \beta_k Covariate_k + \alpha_1 \sigma_j + \alpha_2 \mu_j$$



We use a separate parameter vector ($\alpha$) for the latent $\sigma_j$s and $\mu_j$s to distinguish these latent (i.e., estimated) variables from other predictors or covariates that are part of the observed data. A conceptual diagram of the model is shown in Figure 1. Although the model has two stages, it is estimated simultaneously in one Bayesian model.

[*Insert Figure 1 about here*]

One criticism of the *ISD* approach for estimating individual variability is that it does not take the order of observations into account (e.g., Deboeck et al., 2009; Jahng et al., 2008), and thus mean change over time (e.g., a linear increase) is inevitably included in the estimate of variability. Our method overcomes this limitation by nesting the estimate of variability in a multilevel model, and if there are systematic trends in the data, time or other relevant variables can be added as predictors, and random slopes are also allowed. This gives the flexibility to explicitly model trends in the data, or if it is desired, to leave variability associated with trends in the intra-individual variability estimate. If the functional form of trends is unknown, flexible time trends can be estimated in the proposed model using Gaussian processes (for an introduction, see Rasmussen & Williams, 2006). Further, incorporating detrending into the model rather than performing it in a separate step has the advantage that uncertainty associated with the detrending will be appropriately incorporated into all estimates.

**Software Implementation: VARIAN**

We developed an open source R package for **vari**ability **an**alysis, VARIAN, which is available online at https://github.com/ElkhartGroup/VARIAN, to allow researchers to easily implement the proposed Bayesian variability model. To estimate the models, the VARIAN package links to Stan (Stan Development Team, 2014a), which is a general purpose programming language for Bayesian inference using Markov Chain Monte Carlo (MCMC) and



sampling using the No-U-Turn Sampler which is an extension of Hamiltonian Monte Carlo

(Hoffman & Gelman, 2014). Hamiltonian Monte Carlo requires manual tuning, but is less

sensitive to correlated parameters than Metropolis or Gibbs sampling.  The No-U-Turn Sampler

extends Hamiltonian Monte Carlo by using an algorithm to determine the tuning parameters, so

that manual tuning is not required. At present VARIAN allows only variability of continuous,

normally distributed variables to be used as mediators and outcomes.

        The default priors in the VARIAN package will be weakly informative as long as the

standard deviation of variables is approximately 10 or less.  Means and regression coefficients

use a normal prior with mean zero and standard deviation of 1,000.  The scale and shape

parameters for the gamma distribution and the residual variance from the second stage outcome

use half-Cauchy priors, which has been recommended as a better weakly-informative prior than

either the uniform or inverse-gamma families for variance components (Gelman, 2006; Polson &

Scott, 2012), specifically with location and scale parameters of zero and ten, respectively.

Depending on the scale of the data, it may be necessary to specify alternate parameters for the

priors or to rescale the data in order for the default priors to be weakly informative.

        Convergence can be checked by calculating the percent scale reduction factors (PSRFs;

Brooks & Gelman, 1998; Gelman & Rubin, 1992) for each parameter in the model.  PSRFs, also

referred to as *Rhat*s, estimate the percent scale reduction possible by running the MCMC chains

longer.  A value of one indicates convergence, although typically values sufficiently close to one

are considered indicative of convergence (e.g., < 1.1).  Because each individual estimate of $\sigma_j$

and $\mu_j$ is a parameter, there are many individual PSRFs.  In the VARIAN package, as a

diagnostic plot, we present a histogram of the *Rhat*s to allow easy visual inspection of

convergence for all parameters.



Bayesian inferences uses Bayes' rule to calculate the posterior distribution of the parameters, *p(parameters | data, prior)*, which is the probability of the parameter(s), conditioned on the data and the prior distribution (Gelman et al., 2014). MCMC draws samples from the posterior probability distribution of the parameters. Point estimates are obtained by summarizing the MCMC samples from the posterior distribution, such as by calculating the mean or median of the samples; uncertainty can be characterized by calculating the standard deviation, or presenting the percentiles (e.g., the $2.5^{th}$ and $97.5^{th}$ percentiles for a 95% confidence interval; CI) of the MCMC samples. If *p*-values are desired, two-tailed empirical p-values can be calculated as two times the smaller of the proportion of samples falling above or below zero, that is: *2 \* min(prop(θ ≤ 0), prop(θ > 0))*.

Bayesian inference relies on summarizing the posterior distribution. Therefore, in order to have stable summaries, it is important to have an adequate posterior sample size. However, if there is high autocorrelation in the samples, many posterior samples may not sufficiently characterize the entire posterior parameter distribution. Stan returns an estimate of the effective sample size for each parameter adjusted for autocorrelation estimates based on variograms and the multi-chain variance (Stan Development Team, 2014b). As another diagnostic plot, the VARIAN package presents the effective posterior sample size for each individual parameter. If the estimated posterior effective sample size (i.e., "n_eff") is insufficient, it may indicate that additional iterations are required, or that some other approach such as rescaling the data, simplifying the model, or using stronger priors may be required.

Before we demonstrate how to use the VARIAN package with actual data, we present a Monte Carlo simulation study that compares the performance of the proposed Bayesian model to a model that calculates the *ISDs* and enters these into a multiple linear regression, referred to as



the *ISD* Model.  We have focused specifically on small numbers of repeated measures and small sample sizes because (a) longitudinal data sets are more often than not limited in the number of participants as well as number of replications and because we aimed to develop a model that can be applied in such contexts, and (b) previous methods developed by Hedeker et al. (2008) and Wang et al. (2012) both noted limitations handling smaller numbers of repeated measures and smaller sample sizes.

## Simulation Study

### Simulation Description

For the simulation, the following four factors were varied, resulting in a 2 x 2 x 2 x 2 factorial design with 16 distinct conditions: (1) he number of repeated measures, $k = \{5, 14\}$; this represents the number of observations that may be available from a longitudinal study (5) or a two-week daily assessment study (14); (2)  the sample size, $N = \{80, 250\}$; (3)  the effect size of the relationships between the intra-individual variability estimate and the outcome, standardized $\alpha_1 = \{.2, .5\}$; and (4)  the individual differences in variability by drawing from either $\Gamma(1, .25)$ or $\Gamma(4, 1)$;both have a mean of four, but their standard deviations are four and two, respectively, representing "high" and "low" variability conditions (see Figure 2).

*[Insert Figure 2 about here]*

For the simulation, we refer to the outcome on which intra-individual variability is calculated as *V* and the outcome that is predicted from variability *Y*.  The data were simulated in four steps:

1. *N* individual means were drawn for *V*, with mean zero and standard deviation one: $\mu_j \sim N(0, 1)$.  The mean and standard deviation were fixed for all conditions.

2. *N ISD*s were drawn from a gamma distribution, either $\Gamma(1, .25)$ or $\Gamma(4, 1)$: $\sigma_j \sim \Gamma(\alpha, \beta)$.



3. *k* observations for each of the *N* subjects were drawn from a normal distribution to create the data for *V*: $V_{ij} \sim N(\mu_j, \sigma_j)$.

4. *N* observations were drawn for *Y* from a normal distribution:

   $Y \sim N(\alpha_1 * \sigma_j + \alpha_2 * \mu_j, 1)$.  The intercept of *Y* was fixed at zero, the residual variance at one.  Standardized $\alpha_1$ was either .2 or .5, and standardized $\alpha_2$ was fixed at .3.  Although the variability of $\mu_j$ was held constant at one, the variability of $\sigma_j$ differed between conditions, therefore, the standardized $\alpha$s were transformed to unstandardized coefficients for purposes of simulation, in order to hold the standardized values constant.

We allowed both the intra-individual variabilities ($\sigma_j$) and the individual means ($\mu_j$) to predict the outcome, *Y*, because variability is often calculated on variables that are known to have a mean effect (e.g., borderline personality patients both have higher overall negative affect and have more affective instability) and in order to estimate the unique effect of variability independent of mean effects.  For each of the 16 conditions, we simulated 500 datasets.  The *ISD* Model was analyzed by first calculating the raw score *ISD*s and individual means, and entering these into a multiple regression using the "lm" function in R.

**Bayesian Model Procedures**

For analysis, we ran four chains with independent random seeds.  In all conditions, each of the four chains had 500 burn-in iterations, which were not used for posterior inference.  Based on pilot simulations, we varied the degree of thinning for each condition.  For the "high" variability condition, $\Gamma(1, .25)$, we used a thin of four when $k = 5$ and a thin of two when $k = 14$.  For the "low" variability condition, $\Gamma(4, 1)$, we used a thin of 10 throughout.  We also varied the total number of iterations so that inference from the posterior after thinning would be based on



1,000 MCMC samples in all conditions. These final 1,000 samples were based on combining the 250 samples (after burn-in and thinning) from each of the four independent chains. Start values based on means, *ISD*s, and linear regressions are provided by default by the VARIAN package to speed convergence.

Model convergence was determined as *Rhat*s < 1.1 for all parameters. In practice, if a model converged, but had a low effective posterior sample size, one could simply run the MCMC chains longer until the desired posterior sample size was obtained. However, for the purposes of this simulation, we excluded models where the effective posterior sample size for the focal parameter, $\alpha_1$, was < 200.

**Simulation Outcomes**

For this simulation we focused on the regression parameters $\alpha_1$ and $\alpha_2$ representing the effects of the intra-individual variability and individual mean on the outcome, *Y*. Percent relative bias was used to assess accuracy and was calculated as: $(\hat{\theta} - \theta)/\theta \ x \ 100$, where $\hat{\theta}$ represents the estimated parameter value and $\theta$ represents the true value. The percent relative bias represents the ratio of the bias to the true value and zero bias is optimal. For example a 20 would indicate that the estimate was 20% larger than the true parameter value. Following Hoogland and Boomsma (1998), we consider ± 5% bias acceptable. Coverage assesses how well the model captures the uncertainty of the parameter estimate and was examined as the proportion of CIs from the simulations that included the true parameter value. In this study, 95% CIs were used, so empirical coverage rates of .95 indicate the model performs as expected, with lower coverage rates indicating the CIs are too small (liberal) and higher coverage rates indicating the CIs are too large (conservative). Finally, empirical power was examined as the proportion of 95% CIs that did not include 0. The 95% level was used because a 5% Type I error rate is the most commonly



used in psychology.  For the Bayesian analysis, the mean of the posterior distribution was used as the parameter estimate, and the 2.5[th] and 97.5[th] percentiles were used for the CI.  For the *ISD* Model, the coefficient and normal theory CI were used.

**Results**

The *ISD* Model converged for all conditions and simulations. For the Bayesian analysis, convergence rates and whether there was a sufficient sample size for the $\alpha_1$ parameter are reported in Table 1.  In general, the Bayesian analysis had a high convergence rate; however, in the low variability condition, when $k = 5$ and $N = 80$, the model only converged 48% of the time for both the large and small effect size conditions.  When $k = 5$ and $N = 250$, convergence improved, but still was only 91%.  In the high variability condition, convergence was above 92% for all conditions.  In all conditions, >94% of simulations had >200 effective posterior sample size for the $\alpha_1$ parameter.  For the Bayesian analysis, relative bias, coverage, and power are only reported on those models that converged and had a sufficiently large effective posterior sample size for $\alpha_1$.

*[Insert Table 1 about here]*

**Parameter Estimates**. The average percent relative bias for the parameter estimates across simulations for both the *ISD* Model and the Bayesian model are reported in Table 2.  The *ISD* Model consistently underestimated the true effects of $\sigma_j$ on $Y_j$ ($\alpha_1$) and of $\mu_j$ on $Y_j$ ($\alpha_2$) and yielded unacceptably high bias (greater than ± 5%) regardless of sample size, number of repeated measures, or effect size for the low variability condition.  In the high variability condition, relative bias was acceptably low for $\alpha_1$, only when $k = 14$.  Percent relative bias was larger when $k = 5$ than when $k = 14$.  There was also an effect of variability, with smaller relative bias overall in high than the low variability conditions for $\alpha_1$ but not for $\alpha_2$.  Percent relative bias was similar



across effect size and sample size for both parameters. For $\alpha_2$, percent relative bias was unacceptably high for all conditions.

Turning to the Bayesian model, similar patterns emerged. However, in all conditions, the percent relative bias was closer to zero (i.e., optimal) than in the *ISD* Model. For $\alpha_1$, percent relative bias was acceptably small for all conditions except low variability when $N = 80$ and $k = 5$, and high variability, with a small effect size when $N = 80$ and $k = 5$, although this was only slightly over the threshold at 5.13%. In the low variability condition for $\alpha_2$, relative bias remained above 5% except when $N = 250$ and $k = 14$. For the high variability condition, for $\alpha_2$, relative bias for was acceptable low regardless of sample size, repeated measures, and effect size. For the Bayesian analysis, relative bias appears to respond to both the number of repeated measures and the number of subjects, and in general relative bias is only high when there is low variability and few repeated measures or small sample size. With high variability, even few repeated measures and a small sample size produced acceptable low bias. Finally, when the Bayesian model was biased, it tended to be positively biased, whereas the ISDM tended to be negatively biased.

[*Insert Table 2 about here*]

**Coverage.** Coverage for both the *ISD* Model and Bayesian models are reported in Table 3. For the *ISD* Model, for $\alpha_1$, coverage tended to be lower than the nominal 95%, particularly in the low variability and in the large effect size conditions. There was also an effect of number of repeated measures, with coverage tending to be worse when $k = 5$ than when $k = 14$. For the *ISD* Model for $\alpha_2$, the coverage rates in all conditions were extremely low. As previously noted there was large and systematic negative bias in the *ISD* Model, which combined with overly small CIs



resulted in the upper CI being below the true parameter value for $\alpha_2$ in most conditions. For the

Bayesian analysis, coverage rates were excellent ranging from .92 to .96.

[*Insert Table 3 about here*]

**Power.** Results for empirical power for the *ISD* Model and Bayesian model are reported

in Table 4. Power for the small effect size condition ($\alpha_1 = .20$) was low for both the *ISD* Model

and the Bayesian model for $\alpha_1$ when $N = 80$. Overall there was a small effect of the number of

repeated measures, with higher power for $k = 14$ than for $k = 5$. Both the *ISD* Model and the

Bayesian model had high power to detect a significant effect of $\alpha_1$ in the large effect size

condition ($\alpha_1 = .50$). Power tended to be slightly higher in the Bayesian than in the *ISD* Model.

For $\alpha_2$, power tended to be higher in the Bayesian model than in the *ISD* Model, particularly

when $k = 5$, nevertheless, power remained low in most conditions, except when $N = 250$ and $k = 14$. It is important to note that $\alpha_2$, was fixed at .30 for all simulations, so that the different effect

size conditions apply only to $\alpha_1$.

[*Insert Table 4 about here*]

**Brief Summary.** For the focal parameter $\alpha_1$ (the regression of the outcome on

variability), the Bayesian model yielded near zero relative bias (i.e., good accuracy) as long as

either the variability was high, the number of repeated measures was high ($k = 14$), or the sample

size was large ($N = 250$). Thus using the proposed Bayesian model, researchers may conduct

variability analyses even with relatively small samples and few repeated measures as long as

there are large enough individual differences in variability. In addition, for all conditions, the

coverage rates of the Bayesian model were excellent, indicating that even in those cases where

the estimate is slightly biased, the 95% CI includes the true parameter value close to the nominal

95% rate. However, these results cannot be generalized to conditions that were not included in



the simulation, such as when one or more of the distributional assumptions of the model are violated.

Overall, the *ISD* Model exhibited substantial bias and also poor coverage rates. This is consistent with the low reliability of *ISD* Model, which is associated with greater measurement error that attenuates the estimated effect (e.g., Schmidt & Hunter, 1996). Further, it is important to note that although the present simulation only compared the Bayesian model to the *ISD*, other measures of variability that do not account for measurement error (e.g., *RMSSD*, interquartile range, etc.) will also be biased. We suggest that researchers who wish to model variability using methods that do not account for measurement error, such as *ISD*, first calculate the reliability (see Wang & Grimm, 2012) and only proceed if the reliability is sufficiently high.

## Two Empirical Examples

To illustrate an application of the proposed method and to demonstrate how the VARIAN package can be used to conduct a variability analysis, we used data from a study of sleep and mood in 146 adolescents with an average age of 16.18 years (Bei et al., 2014). During a two-week vacation period, daily sleep duration was indexed using time in bed (TIB) assessed using actigraphs, watch-like devices that measure sleep and wake patterns based on wrist movement. Adolescents also completed the Pittsburgh Sleep Quality Index (PSQI; Buysse, Reynolds, Monk, Berman, & Kupfer, 1989), which was used to create a subjective sleep quality (SSQ) composite score as previously described (Bei, Wiley, Allen, & Trinder, in press), with higher scores indicating worse subjective sleep quality. Adolescents also filled out the Center for Epidemiologic Studies Depression Scale (CES-D; Radloff, 1977), a commonly used, 20 item measure of depressive symptoms. The sample consisted of all participants who completed the questionnaires and at least one day of actigraphy data. The analyzable sample consisted of 140



adolescents, excluding 2 participants (1%) with missing questionnaire data and 4 participants (3%) who did not provide any actigraphy data. The number of days with available actigraphically derived TIB data ranged from 1 to 14 ($M$ = 13.16, mode = 14).

**Example 1: Analyzing the Association between Subjective Sleep Quality and Individual Average and Variability in Sleep Duration**

We refer to the average TIB as mTIB and to the variability in TIB as vTIB. The model assessed mTIB and vTIB as predictors of SSQ, with sex included as a covariate. In this example we include a brief write up of the statistical methods and illustrate how to conduct diagnostic checks. Appendix A includes the computer code.

The model was run with four independent MCMC chains, 1,000 burn-in iterations per chain, and 16,000 total post burn-in iterations (4,000 per chain), with a thin of 2, yielding a total posterior sample size of 8,000. The diagnostic plots from the model are shown in Figure 3. The upper left panel shows *Rhat*s < 1.1 indicating all parameters converged. The upper right panel shows adequate posterior effective sample size for all parameters, with all > 1,000. Histograms and dot plots with 95% CIs for vTIB (individual variabilities, labeled Est_Sigma) and mTIB (individual means, labeled Est_U) indicate that there are no outliers, and suggest that mTIB is normally distributed (an assumption of the model).

[*Insert Figure 3 about here*]

Figure 4 shows the parameter distributions of $\alpha_1$ and $\alpha_2$, the coefficients of the regression of SSQ on vTIB and mTIB, along with their bivariate distribution, and the empirical p-values.

[*Insert Figure 4 about here*]

Results of the analysis are presented in Table 5. The intercept of TIB indicates that on average adolescents spent approximately 9 hours in bed during the two-week vacation (mTIB=9). There



were individual differences in mean TIB, as indicated by the between person standard deviation of mTIB, $\sigma_\mu$ = 0.70, 95% CI = [0.60, 0.82]. The Gamma rate (8.03) and shape (11.21) parameters indicate that on average, the model estimated individual standard deviation (vTIB) was 11.21/8.03 = 1.4 hours. Figure 3 shows the distribution of vTIB. Turning to SSQ, there was a marginally significant effect of sex, with SSQ being somewhat worse for females, $b$ = .86, $p$ = .06. Both vTIB ($b$ = 2.06, $p$ = .002) and mTIB ($b$ = 1.13, $p$ = .006) had statistically significant and unique associations with SSQ. For each 1 hour increase in vTIB there was a 2.06 units higher SSQ, indicating that more variable TIB was associated with worse subjective sleep quality. Similarly, for each one hour increase in mTIB there was a 1.13 unit higher SSQ score, indicating that longer TIB is associated with worse subjective sleep quality.

[*Insert Table 5 about here*]

In summary, during a two-week vacation, both the *average*, and the *variability* of time in bed were uniquely and significantly associated with subjective sleep quality, after controlling for the effects of sex. Having both high variability and long time in bed was associated with worse perceived sleep quality.

**Example 2: Using Variability in Mediation Analysis: Sleep Quality Mediating the Association between Individual Variability in TIB and Mood**

We next demonstrate how intra- individual variability and mean (in this case of TIB) can be used as independent variables in a mediation analysis. Previous analysis of data collected from the same sample found an indirect effect of sleep onset latency on negative mood via SSQ (Bei et al., in press). We hypothesized that SSQ may also serve as a mediator, between TIB and mood, we tested whether the relationship between vTIB (and mTIB) and depressive symptoms severity is mediated by SSQ (see Figure 5). The same method is followed as in example 1



above, except that intra-individual variability and mean TIB were entered into regressions predicting both the mediator (SSQ) and the outcome (depressive symptom severity measured by CES-D).   In these mediation analyses, we control for sex in both the mediator and outcome regressions.

*[Insert Figure 5 about here]*

Code to conduct these analyses is available in Appendix B.  As before, 4 independent MCMC chains were run, with 1,000 burn-in samples, and 16,000 total samples, with a thin of 2, for a final posterior sample size of 8,000.  Diagnostics (not shown) indicated that the model converged (all Rhat < 1.1) and adequate effect sample size (all effective posterior sample size > 1,000).

Results for this analysis are shown in Table 6.  Controlling for sex, vTIB and mTIB had significant effects on SSQ, such that adolescents with more variable TIB and those with longer average TIB had worse subjective sleep quality (*p-value*s < .01).  Examining depressive symptoms, worse SSQ was associated with significantly higher depressive symptoms (*b* = 1.42, *p* < .001).  The direct effect of vTIB on CES-D was negative (*b* = -5.00, *p* = .016), indicating that after controlling for sex, SSQ, and average TIB, higher vTIB was associated with lower depressive symptoms.  There was no significant direct association of mTIB with CES-D. Finally, we tested whether vTIB and mTIB had significant indirect relations with CES-D through SSQ.  Mediation was tested by calculating the distribution of the product of coefficients.  The indirect associations of vTIB with CES-D (*indirect effect* = 3.10, 95% CI = [0.90, 5.94], *p* = .005) and of mTIB with CES-D (*indirect effect* = 1.70, 95% CI = [0.53, 3.17], *p* = .002) were both significant.

*[Insert Table 6 about here]*



**Summary and Discussion**

These results indicate that vTIB has complex associations with CES-D. Through its association with worse SSQ, vTIB was indirectly associated with higher depressive symptoms; however, controlling for SSQ, higher vTIB was associated with lower depressive symptoms. Similarly, through its association with lower SSQ, low mTIB was indirectly associated with higher depressive symptoms; the association of mTIB with depressive symptoms controlling for SSQ was negative and non-significant. Although these data are cross-sectional and do not support causal inference, they do suggest that subjective sleep quality may represent a mechanism linking both the average and the variability of time in bed to depressive symptoms. These intriguing findings suggest that, when variable sleep schedules are not associated with poor subjective sleep, they are not detrimental to affective experiences. Additional research is needed to better understand the causal processes at work. Such future work can compare CESD between good and poor sleepers or experimentally manipulate variability of time in bed.

These examples highlight the utility of models that incorporate both means and intra-individual variability as predictors to glean additional information from repeated measures studies. Using a Bayesian framework also offers flexibility to simultaneously estimate separate regressions in order to test for mediation effects.

**Discussion**

This manuscript described a Bayesian method for estimating intra-individual variability and allows the use of the estimated index of intra-individual variability as a predictor. We presented two empirical examples to demonstrate how the proposed method could be applied in research, including the use of the intra-individual variability as a predictor and mediator.



Using a simulation study, we compared the proposed Bayesian variability model to the *ISD* and found a clear advantage of the proposed Bayesian model to the *ISD* when there were few repeated measures or small individual differences in intra-individual variability. Under these conditions, using *ISD*s as predictors in multiple regression resulted in substantial bias in parameter estimates and poor coverage.  As shown empirically and analytically in previous work (Estabrook et al., 2012; Wang & Grimm, 2012), fewer repeated measures results in lower power as well.  In contrast, the proposed Bayesian variability model yielded minimal percent relative bias ($0 \pm 5\%$) in most conditions, and the bias was lower than using *ISD*s across all conditions. Nevertheless, when there were small individual differences in intra-person variability, only five repeated measures and only 80 subjects, parameter estimates from both methods were biased. Importantly, even under conditions when the parameter estimates were biased, the 95% CIs of the Bayesian variability model, but not the *ISD* model, still performed well, suggesting that even when some bias is present, the CIs are still trustworthy when using the proposed Bayesian variability model.

The present study had several limitations.  At the conceptual level, because we focused on models for few repeated measures (5-14), the proposed Bayesian variability model considers only the magnitude of fluctuations in its assessment of variability, thus addressing a limitation inherent in methods that consider both the magnitude of fluctuations and the temporal dependence that were previously noted by Wang et al. (2012).  Although considering the temporal dependence may be important in some contexts, particularly when there are many repeated measures, it also adds to the complexity of the model and, when there are only a few repeated measures, it may lead to unstable estimation of intra-individual variability.  The simulation study provided preliminary evidence for the performance of the proposed Bayesian



variability model in a limited number of conditions. We only examined performance when the data met all of the assumptions of the model, including data coming from a stationary distribution, the outcomes has a normal distribution and individual standard deviations is based on from a gamma distribution. Whether the model is robust to small or large violations of the distributional assumptions remains to be examined. Nested in a larger Bayesian framework, the proposed method can estimate flexible time trends including standard approaches such as linear or polynomial time effects as well as more flexible data based approaches such as Gaussian Processes. However, it remains unknown how the model performs if no time trend is modelled when a true trend is present. For example, if a linear time trend is modeled, but in reality there is a piecewise trend. Finally, the present study only considered continuous, normally distributed outcomes.

### Recommendations for Selecting Methods for Estimating Intra-individual Variability

The first factor to consider when planning a variability analysis is whether the investigators are interested in intra-individual variability as an outcome or as a predictor. For research questions that require that an estimated index of intra-individual variability be entered into subsequent analysis, it is important to evaluate the reliability of the estimated index of variability. This is particularly important when the number of repeated measures is small. Reliability can be calculated using procedures described by Wang and Grimm (2012). Even if there are many repeated measures, employing an analysis that accounts for measurement error in the individual variability estimates, such as the one we proposed, will be more accurate.

When variability is an outcome, another consideration (in addition to reliability) is whether to focus solely on the magnitude of intra-individual variability, or on both the magnitude and the temporal dependence. Estimating magnitude and temporal dependence requires more



data than estimating magnitude alone; however, when sufficient data are available, examining temporal dependence in addition to magnitude could provide valuable information.  If there is a large number of repeated measures available, and predicting both magnitude and temporal dependence are of interest, we recommend the one step Bayesian method proposed by Wang et al. (2012).  Wang et al. (2012) provide code to conduct the analysis in their appendix using the freely available WinBUGS program.  If the number of repeated measures to examine temporal dependence is small or if an index of variability is not intended to be used as a dependent variable in subsequent analyses, intra-individual variability can be modeled as proposed by Hedeker et al. (2008) and can be applied using the freely available software they developed for mixed effects location scale analysis, MIXREGLS (Hedeker & Nordgren, 2013).  Importantly, the mixed effects location scale analysis is the only one that has been generalized beyond continuous outcomes to ordinal outcomes (Hedeker et al., 2009), but only using PROC NLMIXED in SAS.

When intra-individual variability is examined as a predictor, options are to (a) use estimates of individual variability (e.g., *ISD*) as observed variables in a regression or other model, or (b) use the Bayesian variability model proposed in the current study.  Presently, the method we proposed is developed only for continuous variables and amplitude of fluctuations; nonetheless, this focus also allows for the model to be used with small sample sizes and few repeated measures.  The proposed Bayesian variability model can be employed using the freely available R package, VARIAN that we have developed.  The VARIAN package also implements a mediation model where variability is the independent variable and there is a single mediator, allowing researchers to explore both what variability predicts and potential mechanisms.



When studying variability, power to detect effects depends on a number of factors, such as sample size, the number of repeated measures, and between-person differences in intra-individual variability.  When most participants have similar levels of intra-individual variability, more repeated measures and larger sample sizes are required.  Although presently there is no detailed guide on sample sizes and the number of repeated measures required, our simulation results suggest that when using variability as a predictor, to detect a small effect (standardized $\beta$ = .2, which is equivalent to a correlation of .2), around 250 subjects are required to achieve power of approximately .80, potentially more if there are small individual differences and few repeated measures.  For large effects (standardized $\beta$ = .5, which is equivalent to a correlation of .5), even 80 subjects with only 5 repeated measures yielded power above .90.

**Conclusions and Future Directions**

Recently developed methods allow researchers to investigate variability as an *outcome* and examine factors that may predict intra-individual variability.  The present study expands upon these methods by developing a model for using intra-individual variability as a *predictor* of outcomes.  Monte Carlo simulation demonstrated that the proposed method yields unbiased estimates of the effect of variability on an outcome, with sufficiently large sample size, repeated measures, or individual differences.  The availability of methods and practical software implementation, combined with study designs such as ecological momentary assessment make this is an exciting time for researchers interested in investigating variability, its predictors and its relevance to other outcomes.

Opportunity exists also for methodological work.  Current work is underway to extend the VARIAN package to allow variability to be included in the model simultaneously as a predictor and an outcome.  Variability models also need to be generalized to non-continuous and



non-normally distributed variables.  Theoretical advances or simulations are required to explore the robustness of the current methods to violations of assumptions such as distributions and stationarity.  In addition, research on the number of repeated measures and the sample size required to have sufficient power to detect effects will be paramount in order to provide researchers practical guidance on study design.  Finally, future work is needed to connect variability analysis to existing methods, such as testing mediation and moderation, or estimating variability of latent factors.

Table 1. Percent Convergence for Bayesian Variability Models

| | Low Variability: $\Gamma(4, 1)$ | | | | High Variability: $\Gamma(1, .25)$ | | | |
| | $N = 80$ | | $N = 250$ | | $N = 80$ | | $N = 250$ | |
| $\alpha_I = .20$ | $k = 5$ | $k = 14$ | $k = 5$ | $k = 14$ | $k = 5$ | $k = 14$ | $k = 5$ | $k = 14$ |
|---|---|---|---|---|---|---|---|---|
| Converged | 48.4% | 99.2% | 90.6% | 100% | 97.9% | 99.0% | 96.0% | 92.3% |
| SS | 94.4% | 99.8% | 99.6% | 100% | 97.1% | 99.2% | 96.0% | 99.0% |
| $\alpha_I = .50$ | | | | | | | | |
| Converged | 47.8% | 99.8% | 90.6% | 100% | 97.3% | 98.1% | 98.3% | 93.5% |
| SS | 94.0% | 99.8% | 99.6% | 100% | 97.9% | 98.8% | 98.5% | 99.0% |

*Note.* SS = Sufficient effective posterior sample size (i.e., $\geq 200$).



Table 2. Average Relative Bias x 100 across Simulations

| | Low Variability: $\Gamma(4, 1)$ | | | | High Variability: $\Gamma(1, .25)$ | | | |
|---|---|---|---|---|---|---|---|---|
| | $N = 80$ | | $N = 250$ | | $N = 80$ | | $N = 250$ | |
| | $k = 5$ | $k = 14$ | $k = 5$ | $k = 14$ | $k = 5$ | $k = 14$ | $k = 5$ | $k = 14$ |
| $\alpha_1 = .20$ | $\alpha_1$ (regression of $Y_j$ on $\sigma_j$) | | | | | | | |
| ISDM | -33.32 | -18.98 | -35.89 | -16.75 | -12.54 | -3.18 | -11.48 | -3.87 |
| Bayesian | 16.20 | -2.92 | 1.76 | -1.62 | 5.13 | 2.27 | 4.77 | 0.64 |
| $\alpha_1 = .50$ | | | | | | | | |
| ISDM | -34.51 | -15.90 | -35.64 | -15.37 | -13.72 | -4.12 | -13.74 | -4.80 |
| Bayesian | 10.43 | 0.29 | 2.31 | -0.11 | 3.57 | 1.74 | 2.18 | 0.54 |
| | | | | | | | | |
| $\alpha_1 = .20$ | $\alpha_2$ (regression of $Y_j$ on $\mu_j$) | | | | | | | |
| ISDM | -78.23 | -57.08 | -79.58 | -57.82 | -84.58 | -67.10 | -85.97 | -67.35 |
| Bayesian | 77.80 | 9.72 | 23.45 | 1.45 | -.09 | -1.33 | 0.82 | 0.28 |
| $\alpha_1 = .50$ | | | | | | | | |
| ISDM | -78.21 | -56.94 | -79.69 | -57.92 | -84.57 | -67.03 | -86.01 | -67.45 |
| Bayesian | 75.57 | 8.68 | 17.73 | 1.36 | 0.71 | -1.66 | 0.53 | 0.69 |

*Note*. ISDM = individual standard deviation model using individual standard deviations and individual means as predictors in a multiple regression. Note that in all conditions, $\alpha_2 = .30$.



Table 3. Empirical Coverage of 95% Confidence Intervals

| | Low Variability: $\Gamma(4, 1)$ | | | | High Variability: $\Gamma(1, .25)$ | | | |
| --- | --- | --- | --- | --- | --- | --- | --- | --- |
| | $N = 80$ | | $N = 250$ | | $N = 80$ | | $N = 250$ | |
| | $k = 5$ | $k = 14$ | $k = 5$ | $k = 14$ | $k = 5$ | $k = 14$ | $k = 5$ | $k = 14$ |
| $\alpha_1 = .20$ | $\alpha_1$ (regression of $Y_j$ on $\sigma_j$) | | | | | | | |
| ISDM | .91 | .93 | .71 | .89 | .94 | .94 | .93 | .94 |
| Bayesian | .96 | .95 | .95 | .94 | .95 | .95 | .96 | .95 |
| $\alpha_1 = .50$ | | | | | | | | |
| ISDM | .48 | .84 | .09 | .67 | .83 | .92 | .70 | .90 |
| Bayesian | .94 | .94 | .95 | .93 | .95 | .96 | .96 | .95 |
| | | | | | | | | |
| $\alpha_1 = .20$ | $\alpha_2$ (regression of $Y_j$ on $\mu_j$) | | | | | | | |
| ISDM | .01 | .32 | .00 | .01 | .01 | .15 | .00 | .00 |
| Bayesian | .92 | .95 | .93 | .95 | .94 | .95 | .95 | .96 |
| $\alpha_1 = .50$ | | | | | | | | |
| LM | .01 | .25 | .00 | .01 | .01 | .11 | .00 | .00 |
| Bayesian | .93 | .94 | .94 | .95 | .95 | .96 | .94 | .95 |

*Note.* ISDM = individual standard deviation model using individual standard deviations and individual means as predictors in a multiple regression. Note that in all conditions, $\alpha_2 = .30$.



Table 4. Empirical Power

| | Low Variability: $\Gamma(4, 1)$ | | | | High Variability: $\Gamma(1, .25)$ | | | |
| --- | --- | --- | --- | --- | --- | --- | --- | --- |
| | $N = 80$ | | $N = 250$ | | $N = 80$ | | $N = 250$ | |
| | $k = 5$ | $k = 14$ | $k = 5$ | $k = 14$ | $k = 5$ | $k = 14$ | $k = 5$ | $k = 14$ |
| $\alpha_1 = .20$ | $\alpha_1$ (regression of $Y_j$ on $\sigma_j$) | | | | | | | |
| ISDM | .26 | .35 | .69 | .82 | .35 | .38 | .83 | .87 |
| Bayesian | .29 | .36 | .73 | .83 | .39 | .39 | .85 | .88 |
| $\alpha_1 = .50$ | | | | | | | | |
| IDSM | .96 | .99 | 1.0 | 1.0 | .97 | 1.0 | 1.0 | 1.0 |
| Bayesian | .97 | .99 | 1.0 | 1.0 | .98 | 1.0 | 1.0 | 1.0 |
| | | | | | | | | |
| $\alpha_1 = .20$ | $\alpha_2$ (regression of $Y_j$ on $\mu_j$) | | | | | | | |
| ISDM | .24 | .42 | .59 | .89 | .17 | .34 | .44 | .82 |
| Bayesian | .38 | .49 | .75 | .92 | .41 | .56 | .89 | .97 |
| $\alpha_1 = .50$ | | | | | | | | |
| ISDM | .30 | .50 | .66 | .92 | .21 | .42 | .51 | .88 |
| Bayesian | .42 | .60 | .83 | .97 | .49 | .67 | .95 | .99 |

*Note*. ISDM = individual standard deviation model using individual standard deviations and individual means as predictors in a multiple regression. Note that in all conditions, $\alpha_2 = .30$.



Table 5.  Variability Analysis Results for TIB and SSQ

|  | Estimate | 95% CI | *p*-value |
|---|---|---|---|
| TIB |  |  |  |
|   Intercept | 9.04 | [8.91, 9.17] | < .001 |
|   $\sigma_\mu$ | 0.70 | [0.60, 0.82] | < .001 |
|   Gamma rate | 8.03 | [5.55, 11.39] | < .001 |
|   Gamma shape | 11.21 | [7.87, 15.70] | < .001 |
| SSQ |  |  |  |
|   Intercept | -4.21 | [-6.63, -1.79] | <.001 |
|   sex | 0.86 | [-.04, 1.77] | .062 |
|   vTIB | 2.06 | [0.68, 3.43] | .002 |
|   mTIB | 1.13 | [0.34, 1.91] | .006 |
|   residual | 2.53 | [2.21, 2.90] | < .001 |

*Note.*  TIB = time in bed, $\sigma_\mu$ = standard deviation of the random intercept, SSQ = subjective sleep quality, sex (male = 1, female = 2), vTIB = individual variability in TIB, mTIB = individual mean TIB, residual = residual standard deviation of SSQ.



Table 6.  Variability Analysis Results for Mediation of TIB on CESD through SSQ

| | Estimate | 95% CI | *p*-value |
|---|---|---|---|
| **TIB** | | | |
| Intercept | 9.04 | [8.90, 9.17] | < .001 |
| $\sigma_\mu$ | 0.70 | [0.60, 0.82] | < .001 |
| Gamma rate | 8.07 | [5.63, 11.42] | < .001 |
| Gamma shape | 11.26 | [7.93, 15.84] | < .001 |
| **SSQ** | | | |
| Intercept | -4.33 | [-6.74, -1.77] | .001 |
| sex | 0.84 | [-0.08, 1.75] | .071 |
| vTIB | 2.16 | [0.73, 3.53] | .005 |
| mTIB | 1.19 | [0.42, 1.96] | .002 |
| residual | 2.51 | [2.17, 2.88] | < .001 |
| **CES-D** | | | |
| Intercept | 13.46 | [6.15, 20.63] | < .001 |
| sex | 2.97 | [0.45, 5.54] | .020 |
| SSQ | 1.42 | [0.90, 1.97] | < .001 |
| vTIB | -5.00 | [-9.08, -0.91] | .016 |
| mTIB | -1.76 | [-4.17, 0.55] | .141 |
| residual | 7.14 | [6.29, 8.09] | < .001 |
| **Indirect Effects** | | | |
| vTIB -> SSQ -> CESD | 3.10 | [0.90, 5.94] | .005 |
| mTIB -> SSQ -> CESD | 1.70 | [0.53, 3.17] | .002 |

*Note.*  TIB = time in bed, $\sigma_\mu$ = standard deviation of the random intercept, SSQ = subjective sleep quality, sex (male = 1, female = 2), vTIB = individual variability in TIB, mTIB = individual mean TIB, residual = residual standard deviation of SSQ.



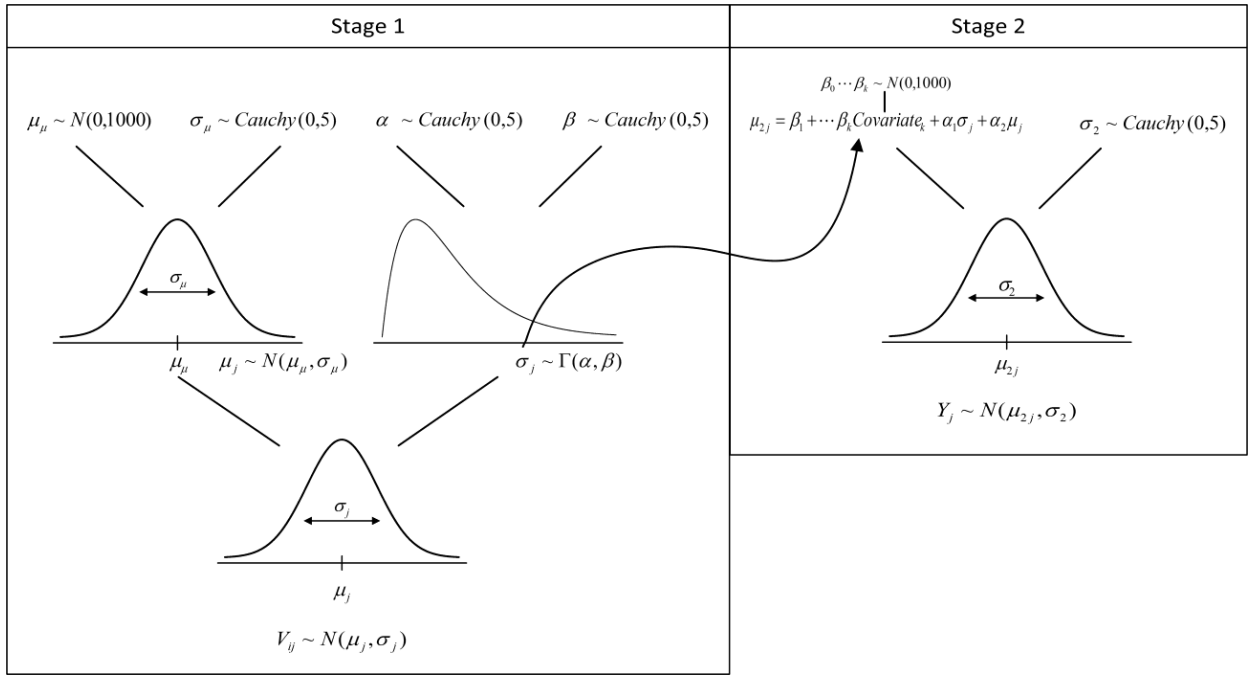

*Figure 1.* Diagram of the Variability Model for prediction.



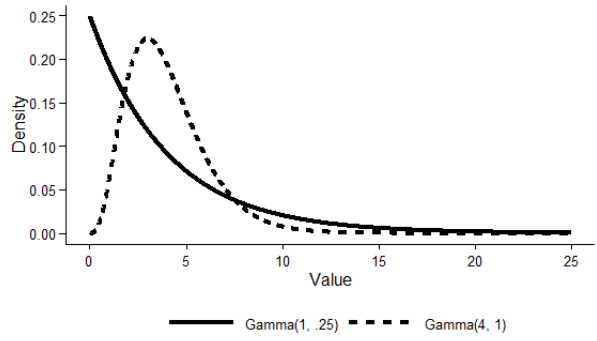

*Figure 2.* Sampling distributions for the intra-individual variabilities, both with means of 4, but

with standard deviations of 2 ("low" condition, $\Gamma(4, 1)$) and 4 ("high" condition, $\Gamma(1, .25)$).



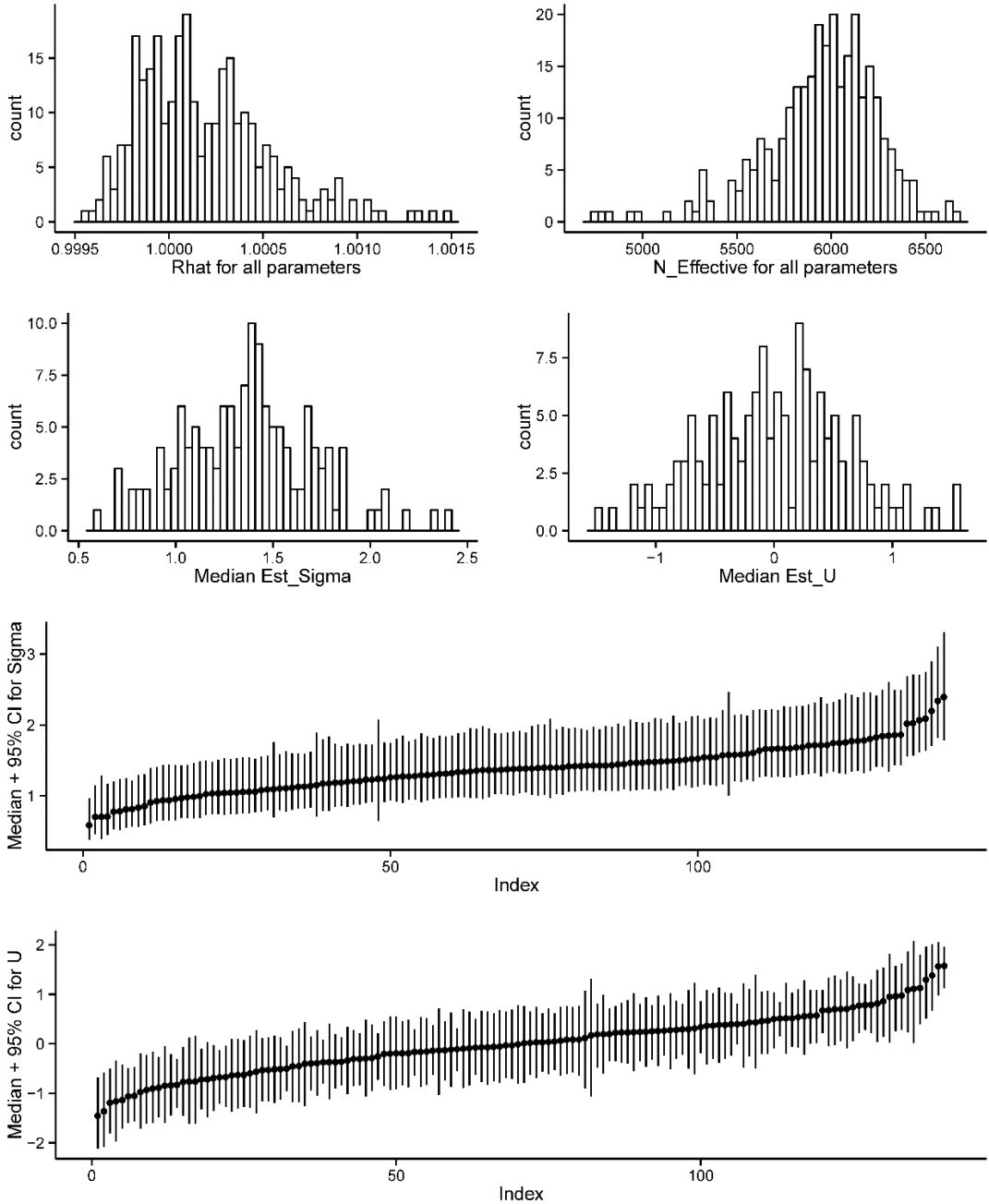

*Figure 3.*  Diagnostic plots for variability analysis showing histograms of percent scale reduction factor (Rhat) and posterior effective sample size (N_Effective) for all parameters.  Histograms of the posterior median for each latent $\sigma_j$s (Sigma) and $\mu_j$s (U) are also shown as well as dotplots of the posterior medians for each subject with 95% credible interval bars.



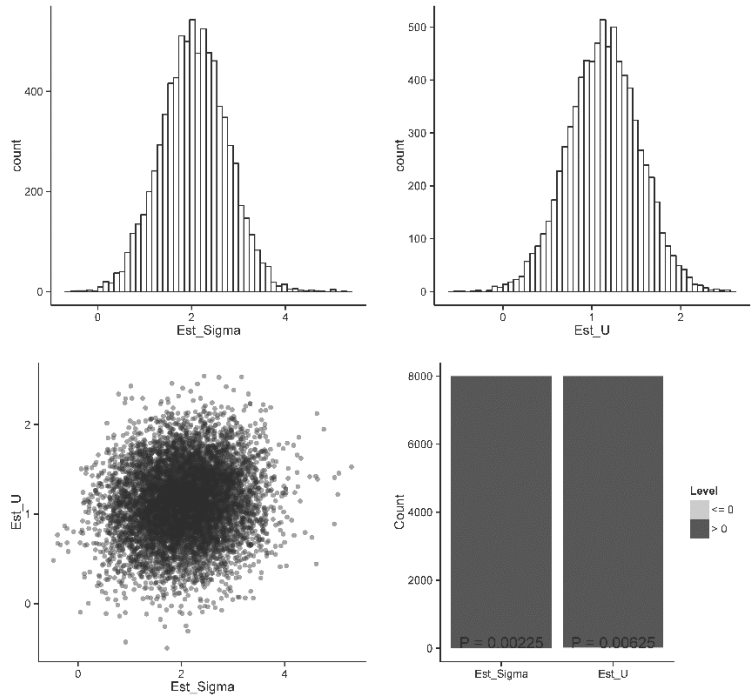

*Figure 4*. Graph of posterior distribution of regression parameters from variability analysis. The top left graph is the posterior distribution of the regression coefficient of SSQ on individual variability of TIB (Est_Sigma). The top right graph is the posterior distribution of the regression coefficient of SSQ on individual mean of TIB (Est_U). The bottom left graph shows the bivariate parameter distribution, indicating that there is low correlation between the two parameters. The bottom right graph is a stacked bar plot of the number of posterior samples falling above and below zero, with the empirical p-values added.



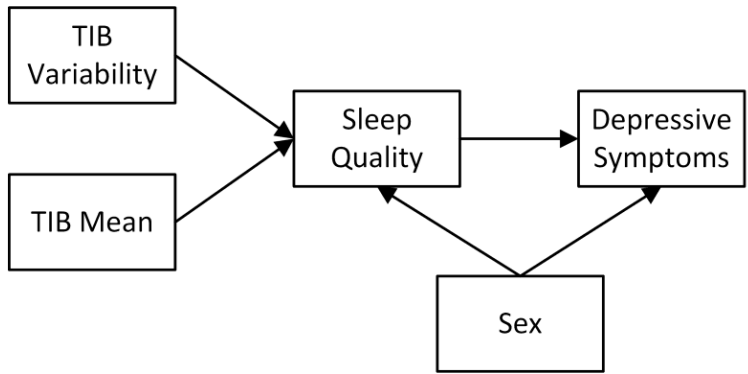

*Figure 5.* Diagram of mediation model.



## Appendix A: R Code for Example 1

```
# load the package
library(varian)

# estimate the model
# SSQ is the outcome (Y)
# Sex is a covariate
# variability of TIB (V) is estimated
m <- varian(SSQ ~ Sex,
 TIB ~ 1 | ID, data = d,
 design = "V -> Y", useU = TRUE,
 totaliter = 16000, warmup = 1000,
 thin = 2, chains = 4)

# create diagnostic plots
vm_diagnostics(m)

# extract MCMC samples
mcmc.samples <- extract(m$results,
  permute = TRUE)

# examine MCMC samples of
# the alpha regression coefficients
vmp_plot(mcmc.samples$Yalpha)

# empirical p-values for all parameters in Table 5
empirical_pvalue(mcmc.samples$VB[, 1])
empirical_pvalue(mcmc.samples$sigma_U)
empirical_pvalue(mcmc.samples$rate)
empirical_pvalue(mcmc.samples$shape)
empirical_pvalue(mcmc.samples$YB[, 1])
empirical_pvalue(mcmc.samples$YB[, 2])
empirical_pvalue(mcmc.samples$Yalpha[, 1])
empirical_pvalue(mcmc.samples$Yalpha[, 2])
empirical_pvalue(mcmc.samples$sigma_Y)
```



**Appendix B: R Code for Example 2**

```
# load the package
library(varian)

# estimate the model
# CESD is the outcome (Y)
# SSQ is the mediator (M)
# Sex is a covariate
# variability in TIB (V) is estimated
m.med <- varian(CESD ~ SSQ + Sex,
 TIB ~ 1 | ID, SSQ ~ Sex,
 data = d, useU=TRUE,
 totaliter = 16000, warmup = 1000,
 thin = 2, chains = 4)

# create diagnostic plots
vm_diagnostics(m)

# extract MCMC samples
med.mcmc.samples <- extract(
 m.med$results, permute = TRUE)

# p-value for the a path
empirical_pvalue(med.mcmc.samples$Malpha[, 1])

# p-value for the a path
empirical_pvalue(med.mcmc.samples$YB[, 2])

# p-value for the indirect effect (a * b)
empirical_pvalue(med.mcmc.samples$Malpha[, 1] *
  med.mcmc.samples$YB[, 2])
```



**Online Supplement**

For the $Y$ intercept, bias is presented instead of relative bias, because the true intercept parameter value was always 0.  Likewise power was not calculated, although coverage is presented.

Table 1. Bias x 100 for $Y$ Intercept

|  | Low Variability: $\Gamma(4, 1)$ | | | | High Variability: $\Gamma(1, .25)$ | | | |
|---|---|---|---|---|---|---|---|---|
|  | $N = 80$ | | $N = 250$ | | $N = 80$ | | $N = 250$ | |
| $\alpha_1 = .20$ | $k = 5$ | $k = 14$ | $k = 5$ | $k = 14$ | $k = 5$ | $k = 14$ | $k = 5$ | $k = 14$ |
| ISDM | 16.28 | 8.90 | 16.78 | 7.86 | 3.30 | 0.55 | 3.66 | 1.39 |
| Bayesian | -5.85 | 1.54 | -0.66 | 0.79 | -1.26 | -0.80 | -0.83 | -0.23 |
| $\alpha_1 = .50$ |  |  |  |  |  |  |  |  |
| ISDM | 47.55 | 21.75 | 48.35 | 20.88 | 11.18 | 3.19 | 11.70 | 4.18 |
| Bayesian | -12.08 | 0.24 | -2.70 | 0.32 | -1.81 | -1.11 | -0.99 | -0.31 |

*Note*. ISDM = individual standard deviation model using individual standard deviations and individual means as predictors in a multiple regression.

Table 2. Empirical Coverage of 95% Confidence Intervals for $Y$ Intercept

|  | Low Variability: $\Gamma(4, 1)$ | | | | High Variability: $\Gamma(1, .25)$ | | | |
|---|---|---|---|---|---|---|---|---|
|  | $N = 80$ | | $N = 250$ | | $N = 80$ | | $N = 250$ | |
| $\alpha_1 = .20$ | $k = 5$ | $k = 14$ | $k = 5$ | $k = 14$ | $k = 5$ | $k = 14$ | $k = 5$ | $k = 14$ |
| ISDM | .90 | .94 | .71 | .90 | .96 | .95 | .91 | .94 |
| Bayesian | .95 | .95 | .96 | .95 | .96 | .95 | .95 | .95 |
| $\alpha_1 = .50$ |  |  |  |  |  |  |  |  |
| ISDM | .46 | .85 | .06 | .70 | .92 | .94 | .76 | .92 |
| Bayesian | .96 | .95 | .95 | .95 | .96 | .95 | .95 | .95 |

*Note*. ISDM = individual standard deviation model using individual standard deviations and individual means as predictors in a multiple regression.